\documentclass[aps,showpacs,twocolumn,superscriptaddress]{revtex4}
\usepackage{bm}
\usepackage{epstopdf}
\usepackage{epsfig}
\usepackage{times}
\usepackage{bbm}
\usepackage{amssymb}
\usepackage{amsmath}
\usepackage{natbib}
\usepackage{color}
\usepackage[latin1]{inputenc}
\usepackage[loose,nice]{units}       

\begin{document}
\title{Envelope induced ionization dynamics beyond the dipole approximation}

\author{Aleksander Skjerlie Simonsen}
\affiliation{Department of Physics and Technology, University of Bergen, NO-7803 Bergen, Norway}

\author{Tor Kjellsson}
\affiliation{Department of Physics, Stockholm University, AlbaNova University Center, SE-106 91 Stockholm, Sweden}

\author{Morten F{\o}rre}
\affiliation{Department of Physics and Technology, University of Bergen, NO-7803 Bergen, Norway}

\author{Eva Lindroth}
\affiliation{Department of Physics, Stockholm University, AlbaNova University Center, SE-106 91 Stockholm, Sweden}

\author{S{\o}lve Selst{\o}}
\affiliation{Faculty of Technology, Art and Design, Oslo and Akershus University College of Applied Sciences, NO-0130 Oslo, Norway}

\begin{abstract}
When atoms and molecules are ionized by laser pulses of finite duration and increasingly high intensities, the validity of the much used dipole approximation, in which the spatial dependence and magnetic component of the external field are neglected, eventually breaks down.
We report that when going beyond the dipole approximation
for the description of atoms exposed to ultraviolet light, the spatial dependence of the pulse shape, the envelope, provides the dominant correction, while the spatial dependence of the carrier is negligible.
We present a first order beyond-dipole correction to the Hamiltonian which accounts exclusively for nondipole effects stemming from the carrier-envelope of the pulse.
We demonstrate by {\it ab initio} calculations for hydrogen that this approximation, which we will refer to as the {\it envelope approximation}, reproduces the full interaction beyond the dipole approximation for absolute and differential observables and proves to be valid for a broad range of high-frequency fields.
This is done both for the Schr{\"o}dinger and the Dirac equation.
Moreover, it is demonstrated that the envelope approximation provides an
interaction-term which gives rise to faster numerical convergence in terms of partial waves compared to its exact counterpart.

\end{abstract}

\pacs{32.80.Fb, 42.50.Hz, 32.80.Rm}

\maketitle

\section{Introduction}

The dipole approximation is frequently applied when studying photoionization of atoms and molecules.
As the approximation consists in assuming the external 
vector potential to be purely homogeneous,
it disregards both the magnetic component and any interaction with higher multi-poles of the electric field completely.
Clearly this becomes too crude an approximation at higher intensities.
Indeed, experimental observations of laser-matter interactions in which the magnetic component of the laser plays a significant role are seen, see, e.g., Refs.~\cite{Moore1999,Smeenk2011}.
Although experimental manifestation of nondipole effects are scarce
for photon energies beyond the optical region, we do expect such observations in the near future as the brilliance of the strongest free-electron lasers continues to increase~\cite{DiPiazza2012}.
Thus, in the theoretical study of photoionization of atoms and molecules in such laser fields, corrections to the dipole approximation must eventually be included.

In the literature we see corrections to the dipole approximation implemented in various ways.
In most of them, the fact that the intensity varies across the laser beam is neglected; one is usually interested in what happens at the laser focus only.
The spatial dependence in the propagation direction of the pulse, however, must be accounted for. In several works this is done by including the full interaction on a numerical grid~\cite{Latinne1994,Kylstra2000,Aldana2001,Forre2005,Forre2006,Madsen2008,Bauke2011}.
Other implementations involve approximations which allow for separating the interaction in spatial and temporal factors.
One way of achieving this is to write the plane wave in terms of exponentials, see, e.g., Refs.~\cite{Mercouris2002,Sel2009,Pindzola2012} or by Taylor expansions~\cite{Bugacov1993,Chirila2002,Meharg2005,Bachau2014,Simonsen2015}.
The latter typically involves the first order correction only.
In a non-relativistic treatment, this may be motivated by the fact that higher order corrections, being proportional to $c^{-2}$ or higher powers of $1/c$, only come into play for field intensities which would necessitate a relativistic treatment.

Accounting for the spatial dependence of the field in the propagation direction, we take the vector potential ${\bf A}$ to depend only on the variable
\begin{equation}
\label{EtaDef}
\eta \equiv t - {\bf k} \cdot {\bf r}/ \omega \quad ,
\end{equation}
where $\omega$ is the central angular frequency of the laser and ${\bf k}$ is the wave vector. These quantities follow the usual dispersion relation $\omega = kc$. Moreover, we will take the polarization direction to be orthogonal to the propagation direction $\hat{\bf k}$, i.e., the Coulomb gauge restriction, $\nabla \cdot {\bf A} = 0$, is imposed.
In addition, we will assume linear polarization.

The vector potential is separated into one factor for the carrier wave and one function providing the pulse envelope profile,
\begin{equation}
\label{Aseparation}
A(\eta) = \frac{E_0}{\omega} f(\eta) \sin(\omega \eta + \varphi) \quad .
\end{equation}
Here $E_0$ is the peak electric field strength and $\varphi$~represents the carrier envelope phase.
The envelope function $f(\eta)$ is frequently chosen to be the square of a trigonometric function over half a cycle,
\begin{equation}
\label{SinSq}
f(\eta) = \left\{ \begin{array}{lc} \sin^2 \frac{\pi \eta}{T} , & 0 \leq \eta \leq T \\
0 , & \text{otherwise} \end{array} \right. \quad .
\end{equation}
 Another common choice is a Gaussian
\begin{equation}
\label{Gauss}
f(\eta) = \exp \left[- 4 \ln 2 \left( \frac{\eta}{T} \right)^2 \right] \quad ,
\end{equation}
where $T$ in the latter case is the full width at half maximum pulse duration.
A third frequently used way of modeling the pulse is a trapezoidal shape, i.e., the pulse has a linear ramp on/off of relatively short duration at each end while held constant in between.

In Ref.~\cite{Forre2014} it was shown that, for relatively short XUV laser fields, a first
order spatial expansion of the vector potential indeed accounts for practically all corrections
to the dipole approximation when it comes to the ionization probability.
Moreover, the photoelectron spectrum as well as angular-resolved probability distributions
were also well accounted for by this comparatively simple approach.

In this work we will present an approximate form of the nondipole interaction Hamiltonian which turns out to be highly accurate.
In effect, the approximation, which we shall coin the {\it envelope approximation},
neglects the spatial dependence of the carrier-part of the laser field altogether.
To some extent, this may contradict intuition, and
indeed, in literature we do find works in which the spatial dependence of the
envelope is neglected instead~\cite{Mercouris2002,Meharg2005}.
Other theoretical
works beyond the dipole approximation do not involve any envelope at all \cite{Pindzola2012}.
While these approaches may be adequate in certain cases, we demonstrate that
in the general case the opposite is true in the high frequency regime;
the spatial expansion of the envelope alone provides the dominant correction to the dipole
approximation -- in particular for short pulses.
Furthermore, we show that the same conclusion holds for both
the Schr\"{o}dinger and the Dirac equation.

We explain the accuracy and predictive power of the envelope approximation by analyzing the Hamiltonian in a generalized velocity-gauge form, i.e., the beyond-dipole correction is included in a form similar to the (velocity-gauge) dipole interaction.
Two nondipole interaction terms are identified: one ``envelope term'' which represents the time-average of the induced motion along the laser direction of propagation and one ``carrier term'' which causes rapid oscillations.
We argue that the net contribution of the latter term vanishes, thus permitting a scheme in which only the former term is retained.

The envelope approximation is also attractive from a computational point of view, both in the non-relativistic and the relativistic implementations. Numerical simulations are subject to milder convergence criteria within the envelope approximation compared to simulations including the full interaction.
In the following section the approximation is outlined in detail, and its applicability is justified.
Furthermore, in Sec.~\ref{Results}, its adequacy is demonstrated by explicit numerical simulations.

\section{Theory and implementation}
\label{Theory}

Both the non-relativistic Schr{\"o}dinger equation and its relativistic counterpart, the Dirac equation, have the generic form
\begin{equation}
\label{TDSDE}
i \hbar \frac{\mathrm{\partial}}{\mathrm{\partial t}} \Psi = H \Psi \quad ,
\end{equation}
where the wave function $\Psi$ is a four-spinor in the latter case, and $H$ is the Hamiltonian.

\subsection{The Schr{\"o}dinger equation}
For the case of the Schr{\"o}dinger equation of a hydrogen atom exposed to the external field ${\bf A}$, the Hamiltonian $H$ reads
\begin{align}
\label{SchrodH}
H & = \frac{1}{2m} \left[ {\bf p}+ e{\bf A}(\eta) \right]^2 + V(r)  \\
\nonumber
&=H_0 + \frac{e}{m} {\bf p} \cdot {\bf A} + \frac{e^2}{2m} A^2 \quad ,
\end{align}
where $H_0=\frac{p^2}{2m} + V(r)$ is the field-free Hamiltonian, and
$V$ is the Coulomb potential. The last term in Eq.~(\ref{SchrodH}), which is proportional to the square of
the vector potential, is commonly referred to as the {\it diamagnetic} term. It is well
known that this term alone provides the dominant nondipole correction; the spatial
dependence of $(e/m) \; {\bf p} \cdot {\bf A}(\eta)$ has been found to be insignificant in this context \cite{Kylstra2000,Aldana2001,Meharg2005,Forre2014}.

If we resort to a first order expansion of the vector potential around $\eta=t$, and
keep only the first order contribution to the diamagnetic term, the corresponding
Schr{\"o}dinger Hamiltonian takes the form \cite{Forre2014}
\begin{equation}
\label{SchrodH1st}
H_\mathrm{1st}  =
H_\mathrm{dip}(t) - \frac{e^2}{mc} \, x \,A(t) A'(t)
\quad ,
\end{equation}
where $H_\mathrm{dip}$ is the
Schr{\"o}dinger Hamiltonian including only the
dipole part of the interaction with the external electromagnetic field, which in the velocity-gauge form reads
\begin{equation}
\label{SchrodHDip}
H_\mathrm{dip}(t)  = H_0+ \frac{e}{m}  p_z A(t)
\quad .
\end{equation}
We have here taken the $z$-polarized external field to
propagate along the $x$-axis.
Moreover, the purely time-dependent zeroth order term in the diamagnetic
interaction has been removed by a rather trivial phase transformation.

In order to derive the envelope approximation and the corresponding form of the Hamiltonian,
we consider the nondipole interaction within the {\it propagation gauge} form \cite{Forre2015}.
In the particularly simple case of a first order expansion of the field, the transformation \begin{equation}
\label{PropGaugeTrans}
\Psi_\mathrm{PG} = U \Psi, \hspace{0.5 cm} U = \exp \left(-i\frac{e^2}{2mc\hbar} x [A(t)]^2\right)
\end{equation}
transforms the initial Hamiltonian Eq.~(\ref{SchrodH1st}) into \cite{Kim1999,Ryabikin2000,Aldana2001,Emelin2014,Forre2015}
\begin{align}
\nonumber
 H_\mathrm{PG} & = UH_\mathrm{1st}U^{\dagger} + i \hbar \dot{U} U^{\dagger} \\
& = H_0 + \frac{e}{m} A(t) p_z + \frac{e^2}{2m^2c} [A(t)]^2 p_x \quad .
\label{HpressureFull}
\end{align}
The advantage here is that the beyond-dipole term, $e^2/ (2 m^2 c) [A(t)]^2 p_x$, and the velocity
gauge dipole-interaction term, $(e/m) A(t) p_z$,~are of similar forms and can be compared on an equal footing.
The two terms
can be associated with the corresponding
classical motion of a free particle subjected to
the electromagnetic field~\cite{Forre2015}.
The dipole field drives the charged particle back and forth along the direction of polarization
whereas the nondipole field induces a perpendicular motion in the propagation direction,
associated with the (homogeneous) magnetic field.
Since $A^2$ is strictly non-negative, the
interaction along the propagation axis is very different from the dipole-driven dynamics
in that the induced momentum in the former case points in one direction only,
giving rise to a non-negative velocity component in the laser propagation direction.
As such, the nondipole field associated with the radiation pressure of the electromagnetic field
mimics a so-called non-zero displacement pulse~\cite{Ivanov_2014},
which in the present case takes the form
\begin{equation}
\label{AsqHomoField}
\frac{e^2}{2m^2c} [A(t)]^2 = \frac{1}{4c} \left( \frac{eE_0}{m\omega}\right)^2 [f(t)]^2
\left[1-\cos(2 \omega t + 2 \varphi) \right] \quad .
\end{equation}
The nondipole field can further be decomposed into two components, one which carries the non-zero displacement
characteristics of the pulse, the envelope component, and one component
that in itself represents an ordinary zero-displacement pulse. With this decomposition
the propagation gauge Hamiltonian~(\ref{HpressureFull}) becomes
\begin{align}
\label{eq:HamPGFull}
H_\mathrm{PG} = H_0 &+ \left(\frac{eE_0}{m\omega} \right) f(t) \sin(\omega t + \varphi) p_z \\
 &+ \frac{1}{4c} \left(\frac{eE_0}{m\omega} \right)^2 f^2(t) p_x  \nonumber \\
 &- \frac{1}{4c} \left(\frac{eE_0}{m\omega} \right)^2 f^2(t) \cos(2 \omega t + 2 \varphi) p_x \quad .\nonumber
 \end{align}
The first term beyond the dipole approximation, i.e., the envelope component which is proportional to $f^2(t) p_x$,
imparts a net momentum transfer to the particle
in the propagation
direction throughout the laser pulse, while the last term,
proportional to $f^2(t) \cos(2 \omega t + 2 \varphi)p_x$,
only induces high-frequency oscillations around this mean.
Now, since these oscillations are
superimposed on a net shift which changes on a comparatively
slow time scale, their contribution is expected to vanish.
This is in fact the situation when $f^2(t)$ varies
slowly compared to $\pi/\omega$, i.e., when the pulse extends over several optical cycles.
Integrated over the duration of the laser pulse it is only the non-oscillating component,
i.e., the envelope component,
that will induce a net displacement of the particle, effectively
providing the main contribution beyond the dipole approximation.
Therefore, from a physical point of view, and in the case of high-intensity and high-frequency fields, the important
nondipole correction to the laser-atom dynamics stems from the intimate interplay between
the dipole field and the
displacement of the wave packet in the propagation direction.

Correspondingly, the envelope approximation is obtained by neglecting the last term in Eq.~(\ref{eq:HamPGFull}):
\begin{align}
\label{eq:HamPGEnv}
H_\mathrm{PG} \rightarrow H_\mathrm{PG}^\mathrm{Env} = H_0 &+ \left(\frac{eE_0}{m\omega} \right) f(t) \sin(\omega t + \varphi) p_z \\
 &+ \frac{1}{4c} \left(\frac{eE_0}{m\omega} \right)^2 f^2(t) p_x \nonumber \quad .
\end{align}
Going from Eq.~(\ref{eq:HamPGFull}) to Eq.~(\ref{eq:HamPGEnv}) effectively constitutes a time-averaging over the carrier-wave in the nondipole operator, i.e., we have substituted the nondipole correction with its zeroth order Floquet component in terms of the carrier. In this context it is worth noting that also for the dipole field within the Kramers-Henneberger frame,
truncated Floquet expansions of the carrier field have proven useful~\cite{Toyota2007,Toyota2015}.

While the non-zero displacement component of the first beyond-dipole term in Eq.~(\ref{eq:HamPGFull}) dictates the corresponding dynamics, c.f., Ref.~\cite{Dimitrovski2009}, perturbation theory applies to the last term. Although, as we have argued, this term would contribute marginally to the total ionization probability, one might expect it to provide a peak near the ``resonance energy'' at $2 \hbar \omega - I_p$, with $I_p$ being the ionization potential, in the energy differential ionization probability. However, due to the presence of the much stronger dipole field, the initial state becomes strongly field dressed, and the (neglected) interaction would ionize the electron from a state without a well defined energy. Consequently, any such peak is broadened and diminished to such an extent that it in effect vanishes.

Transforming the envelope approximation
Hamiltonian~(\ref{eq:HamPGEnv}) back to the original velocity gauge formulation is
performed with the inverse of transformation~(\ref{PropGaugeTrans}) subject to the envelope approximation, \begin{equation}
\label{Inverstrans}
\Psi = \exp \left( i\frac{e^2 E_0^2}{4mc\hbar \omega^2} x \left[ f(t) \right]^2 \right) \Psi_\mathrm{PG}
\quad ,
\end{equation}
resulting in the following effective (envelope) Hamiltonian:
\begin{equation}
\label{EnvelopeApprox}
H_\mathrm{env} = H_\mathrm{dip}(t) - \frac{e^2}{2mc} \, x \,\left( \frac{E_0}{\omega}\right)^2 f(t) f'(t) \quad .
\end{equation}
Equation~(\ref{EnvelopeApprox}) constitutes one of the main results
of this paper.
We will demonstrate by numerical simulations that it indeed leads to essentially the same final state as
the Hamiltonian including the full first order beyond-dipole correction given by Eq.~(\ref{SchrodH1st}).

\subsection{The Dirac equation}
\label{SubSectTDDE}

For external fields strong enough to potentially accelerate the electron to velocities comparable to the speed of light, the non-relativistic Schr{\"o}dinger Hamiltonian has to be replaced with the corresponding Dirac Hamiltonian:
\begin{equation}
\label{DiracH}
H_\mathrm{D} = c \boldsymbol{\alpha} \cdot \left[ {\bf p} + e {\bf A}(\eta) \right] + V(r) \mathbbm{1}_4 + m c^2 \beta
\quad ,
\end{equation}
with
\begin{equation}
\label{AlphaMat}
\boldsymbol{\alpha}
= \left( \begin{array}{cc} 0 &
\boldsymbol{\sigma}
\\
\boldsymbol{\sigma}
&  0 \end{array}\right) \quad ,
\end{equation}
where $\boldsymbol{\sigma}=(\sigma_x, \sigma_y, \sigma_z)$ are the Pauli matrices and
\begin{equation}
\label{BetaMat}
\beta = \left( \begin{array}{cc} \mathbbm{1}_2 & 0 \\  0 &  -\mathbbm{1}_2 \end{array} \right) \quad .
\end{equation}
An expansion of the Dirac Hamiltonian to first order while
omitting the spatial dependence of the carrier yields
\begin{align}
\label{DiracH1st}
H_\mathrm{D, env} & = c \boldsymbol{\alpha} \cdot {\bf p} + V(r) + mc^2 \beta + ec A(t) \alpha_z
\\
\nonumber
 &- \frac{e E_0}{\omega} f'(t) \sin(\omega t + \varphi) \alpha_z  x
\quad ,
\end{align}
again with linear polarization along the $z$-axis and propagation along the $x$-axis. Equation (\ref{DiracH1st}) provides the corresponding envelope approximation within the Dirac picture, although the interaction has a slightly different
form.

Considering that the Dirac equation is to account for the same physics as the Schr\"{o}dinger equation with the same physical fields, at least in the non-relativistic regime, we expect the envelope approximation to be valid for the Dirac equation as well in the weakly relativistic regime.
From a more technical point of view, we could also apply the above analysis in the non-relativistic limit via a
modified form \cite{Sel2009} of the well established Foldy-Wouthuysen-like transformation \cite{foldy:50}.
The transformation given in Ref.~\cite{Sel2009} yields, after omission of terms proportional to $c^{-2}$ or smaller, the alternative form of the Dirac Hamiltonian
\begin{align}
\label{FWonHdirac}
H_\mathrm{D, alt} & = c \boldsymbol{\alpha} \cdot {\bf p} + \beta mc^2 + V +  \frac{e}{m} \beta \mathbf{A} \cdot \mathbf{p} +
\frac{e^2}{2m}  \beta A^2 \nonumber \\
 & + \frac{e\hbar}{2m} \beta \boldsymbol{\sigma}\cdot  \left( \nabla \times \mathbf{A} \right) \quad ,
\end{align}
where the two upper components of the four-component wave function no longer couple to the two lower through the external electromagnetic field.
Apart from the spin-dependent term, which is of minor importance here, the interaction terms working on the upper components are the same as in the Schr{\"o}dinger equation, Eq.~(\ref{SchrodH}), and thus the same analysis concerning the importance of different contributions applies.
Naturally, the use of
Eq.~(\ref{DiracH1st}) will result in spatial contributions not only to the diamagnetic term, but also to the fourth
term of
Eq.~(\ref{FWonHdirac}), as well as resulting in a non-vanishing sixth term.
As such, by construction, the relativistic version of the envelope approximation, Eq.~\ref{DiracH1st}, will not be exactly identical to
the corresponding approximate scheme in the Schr\"{o}dinger picture, Eq.~(\ref{EnvelopeApprox}),
-- not even in the non-relativistic limit.
However, since the conclusions from the non-relativistic treatment should still hold, we expect the
contribution from the $A^2$-term to be completely dominating in the weakly relativistic region.

\subsection{Numerical implementation}
The time-dependent Schr{\"o}dinger equation (TDSE) has been solved in two ways.
In the first case, the time-dependent state $\Psi$ is expanded directly on a product basis comprising B-splines in the radial direction and spherical harmonics in the angular coordinates.
Then, the resulting sparse system of ordinary differential equations is propagated in time by using a Crank-Nicolson-like method on a rather large simulation domain, see Ref.~\cite{Forre2014} for details.
In the second method, a spectral basis consisting of the eigenstates of the time-independent part of the Hamiltonian is constructed and complex scaling is used in order to minimize the size of the basis set, c.f., Ref~\cite{Beng2008}, while the time propagation is performed with a Magnus type propagator in a Krylov subspace, using the Arnoldi algorithm.
The Dirac equation is solved in a manner rather analogous to the latter method \footnote{Details on the method and implementation is to be published elsewhere.}, although complications arise due to the negative energy continuum~\cite{Sel2009} inherent in the Dirac Hamiltonian, Eq.~(\ref{DiracH}).

\section{Results and discussion}
\label{Results}

\begin{figure}
  \includegraphics[width=9.3cm]{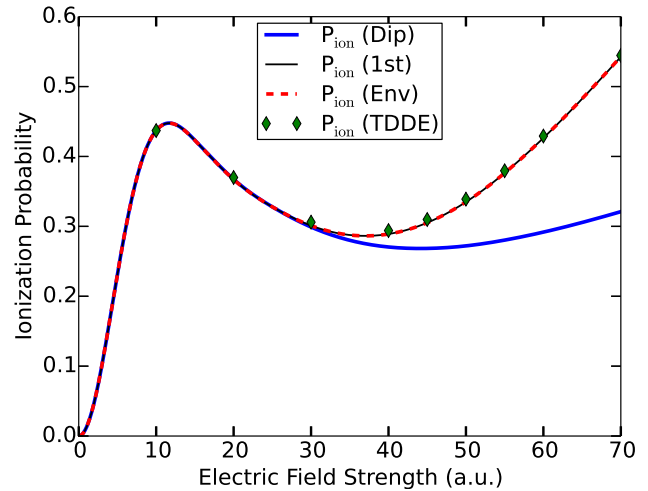}
  \caption{(Color online) The total ionization probability as a function of the laser peak intensity (given in electric field strength).
The hydrogen atom was exposed to a 15-cycle sine-square shaped XUV laser pulse of angular frequency
$\omega = 3.5$~a.u. The results are obtained with four different theoretical frameworks.
The time-dependent Schr\"{o}dinger equation (TDSE) is solved within the dipole approximation, Eq.~(\ref{SchrodHDip}), (solid blue) and for a Hamiltonian including the first order expansion of the full nondipole interaction, Eq.~(\ref{SchrodH1st}), (thin solid black).
The envelope approximation is shown both when applied to the TDSE, Eq.~(\ref{EnvelopeApprox}), (dashed red) and the
time-dependent Dirac equation (TDDE),
Eq.~(\ref{DiracH1st}), (green diamonds).
\label{Fig1}}
\end{figure}

\begin{figure}
  \includegraphics[width=8.8cm]{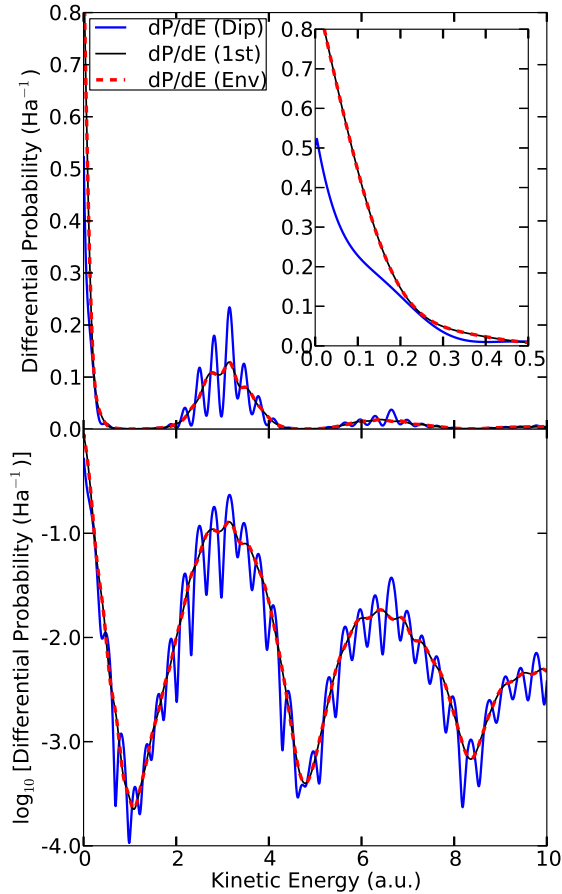}
  \caption{(Color online)
Differential probability distribution for the photoelectron as a function of kinetic energy.
Here, the laser field parameters are the same as in Fig.~\ref{Fig1} with the field strength $E_0 = 45$~a.u. The energy distribution is shown for three different interactions used to model the laser-atom dynamics in the time-dependent Schr\"{o}dinger equation. First, the dipole approximation results are shown in a solid blue line whereas the corresponding predictions by a beyond-dipole interaction to first order is depicted in a thinner solid black line.
Lastly, the results obtained by applying the envelope approximation, i.e., Eq.~(\ref{EnvelopeApprox}), is drawn in a dashed red line.
The energy distributions are shown on absolute (upper) and logarithmic (lower) scale. The inset in the upper panel is a close up on the low energy region.
  \label{Fig2}
}
\end{figure}

\begin{figure}
  \includegraphics[width=6.8cm]{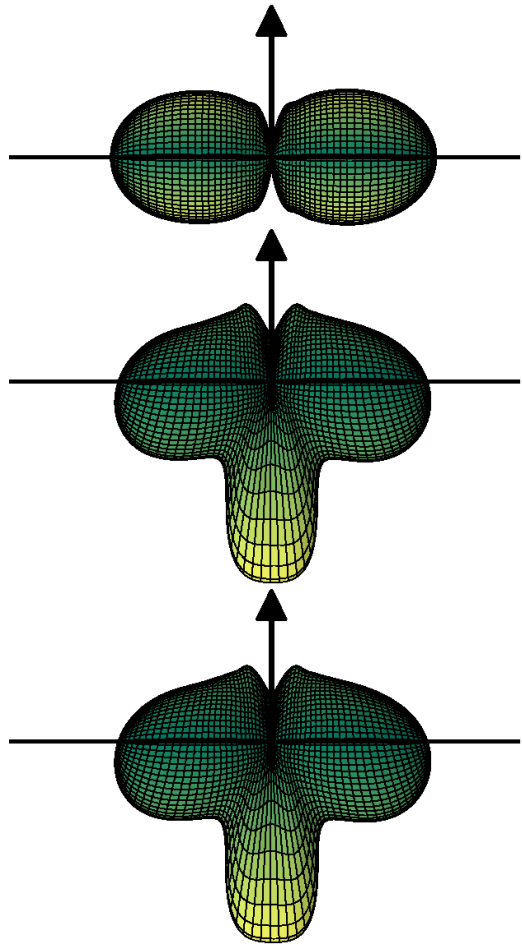}
  \caption{(Color online)
Angular-resolved probability distributions for the final continuum state obtained after the conclusion of the laser pulse used in Fig.~\ref{Fig1} for the field strength $E_0 = 45$~a.u.
Here, the axis of polarization is oriented horizontally whereas the propagation direction is denoted by the vertical black arrows.
The panels correspond to, from the top to bottom, the dipole approximation, Eq.~(\ref{SchrodHDip}), beyond the dipole approximation to the first order, Eq.~(\ref{SchrodH1st}), and the envelope approximation, Eq.~(\ref{EnvelopeApprox}).
  \label{Fig3}
}
\end{figure}

\begin{figure}
  \includegraphics[width=9.8cm]{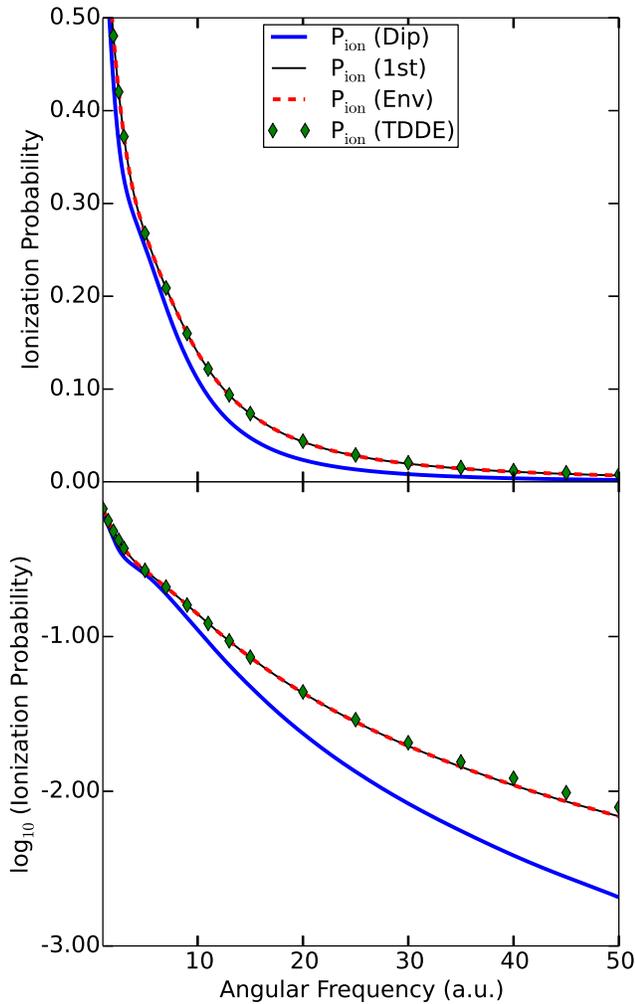}
  \caption{(Color online) The total ionization probability as a function of the laser angular frequency.
The field intensity is adjusted so that the quiver velocity of the free (classical) electron is kept fixed at $10$\% of the speed of light for all the calculations.
The pulse is described by a sine-square envelope function lasting 10 optical cycles.
Three different interactions within the time-dependent Schr\"{o}dinger equation (TDSE) are shown and compared to the results obtained with the time-dependent Dirac equation (green diamonds).
The TDSE Hamiltonians include the dipole approximation, Eq.~(\ref{SchrodHDip}), (solid blue), the dipole interaction and the full first order nondipole correction (thin solid black), Eq.~(\ref{SchrodH1st}), and the envelope approximation applied to the latter (dashed red), Eq.~(\ref{EnvelopeApprox}).
The probabilities are shown on absolute (upper) and logarithmic (lower) scale.
\label{Fig4}
}
\end{figure}

\begin{figure}
 \includegraphics[width = 9.8cm]{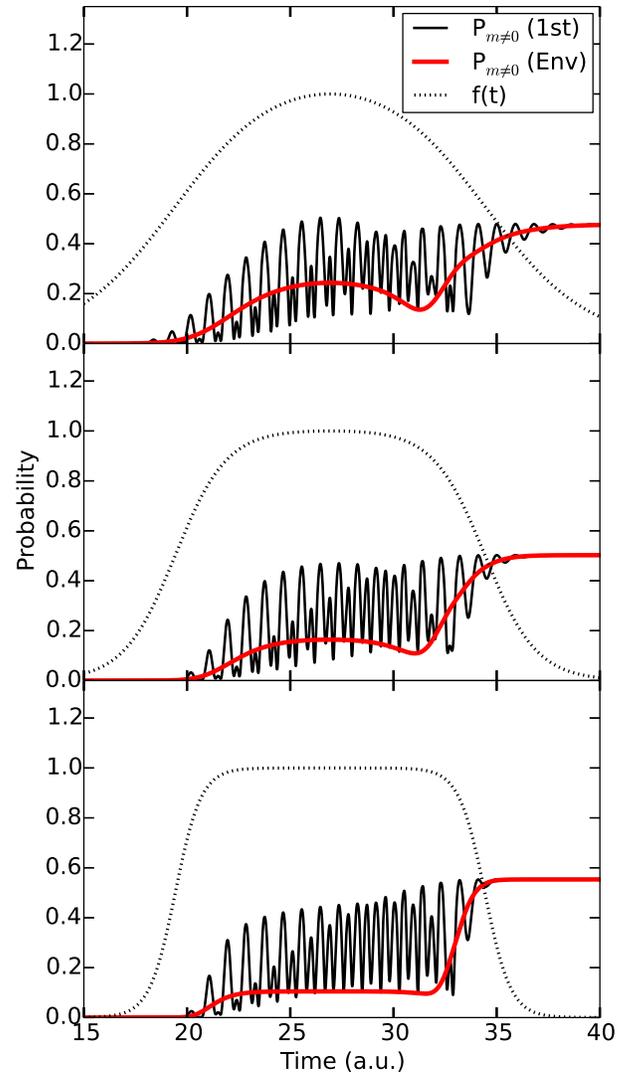}
 \caption{(Color online)
Population in all $m \ne 0$ states as a function of time.
Here, the photon energy is $3.5$~a.u., and the laser intensity corresponds to $E_0 = 45$~a.u. at maximum. The plots show the results obtained with both the full first order nondipole interaction Hamiltonian, i.e., Eq.~(\ref{SchrodH1st}), (thin black lines) and the envelope approximation, Eq.~(\ref{EnvelopeApprox}) (thick red lines).
The dotted black lines depict the shape of the laser pulse, $f(t)$.
Three different envelope functions described by Eq.~(\ref{FermiDiracPuls}) with $\sigma = 0.4$ (top panel), $\sigma = 0.8$ (center panel) and $\sigma = 1.6$ (bottom panel) have been
used to model the laser. In all three cases the pulse duration parameter $T$ corresponds to 7.5 optical cycles.
\label{Fig5} }
\end{figure}

\begin{figure}
 \includegraphics[width = 7.8cm]{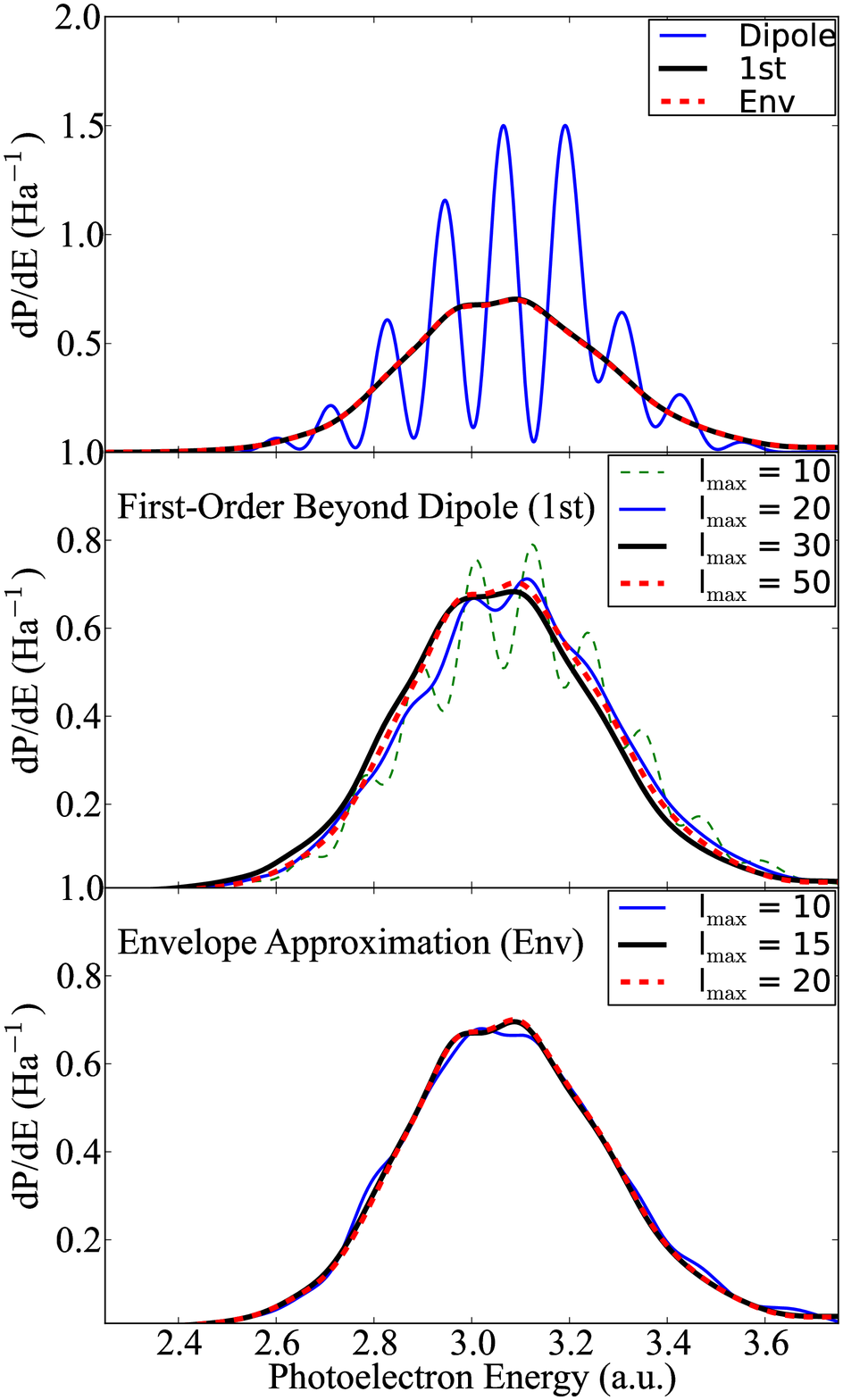}
 \caption{
(Color online)
Energy-distribution of the ejected photoelectron after a 40-cycle laser-atom interaction.
The photon energy and the field intensity are set to $\hbar \omega = 3.5$~a.u. and $E_0 = 45$~a.u., respectively.
The distributions are drawn around the one-photon absorption resonance $E \sim \hbar \omega - I_p$, and the different curves pertain to different values of $l_\mathrm{max}$.
The upper panel shows comparison between fully converged dipole, beyond-dipole (1st) and envelope approximation (Env) results.
The distribution obtained within the dipole approximation features oscillations which vanish when the spatial dependence of the external field is included.
The second and third panels show the results as obtained with the first-order nondipole expansion (1st) and with the envelope approximation (Env) at different values of $l_\mathrm{max}$, respectively.
\label{Fig6} }
\end{figure}

In order to establish the validity of the envelope approximation, we compare results for a hydrogen atom initially prepared in the ground state obtained with the original form of the first order Hamiltonian including corrections beyond the dipole approximation,
$H_\mathrm{1st}$ in
Eq.~(\ref{SchrodH1st}),
and the approximate versions $H_\mathrm{env}$ and $H_\mathrm{D,env}$, in Eqs.~(\ref{EnvelopeApprox}) and (\ref{DiracH1st}), respectively.
The relativistic treatment is partly motivated by the fact that
our calculations are performed in a domain of laser field parameters close to the relativistic regime, i.e., for field intensities so strong that the corresponding classical electron would attain velocities in the order of 10\% of the speed of light.
Moreover, we are interested in confirming that neglecting the spatial dependence of the carrier remains valid also in the relativistic framework (granted that relativistic corrections are small).
The corresponding results obtained with a pure dipole interaction
Hamiltonian, c.f., Eq.~(\ref{SchrodHDip}), are also included in order to demonstrate that the dipole approximation actually breaks down.

Figure~\ref{Fig1} shows the ionization probability as a function of electric field strength for the four different Hamiltonians mentioned above.
The central angular frequency of the field is here $\omega=3.5$~a.u., corresponding to a photon energy of 95~eV, and the pulse profile is described by a sine-square envelope, c.f., Eq.~(\ref{SinSq}), and with a duration of 15 optical cycles, which corresponds to 652~as.
The local maximum of ionization, when the electric field strength is roughly equal to 11~a.u., and the following decrease of ionization for increasing field intensity are predicted by all the interactions considered, and the phenomenon is known as atomic stabilization (see e.g., \cite{Pont1988, Kulander1991, Eberly1993}).
Furthermore, it is clearly seen that the dipole approximation breaks down at around $E_0=30$~a.u.
More importantly, we also see that neglecting the spatial derivative of the carrier
leaves the results virtually
indistinguishable from the full solution; both the approximative Hamiltonians Eqs.~(\ref{EnvelopeApprox}) and (\ref{DiracH1st}) provide the correct (non-relativistic) total ionization probabilities within the Schr\"{o}dinger and Dirac pictures, respectively.

Turning to differential observables, Fig.~\ref{Fig2} shows the energy-differential probability distribution for the ionized electron. Here, the calculations are performed with the same laser pulse parameters as used in Fig.~\ref{Fig1} and for the electric field strength value $E_0 = 45$~a.u. The probability distributions obtained when modeling the dynamics with the three Schr\"{o}dinger Hamiltonians Eqs.~(\ref{SchrodH1st}), (\ref{SchrodHDip}) and (\ref{EnvelopeApprox}) are compared to each other. The figure clearly illustrates that solving the time-dependent Schr\"{o}dinger equation with the Hamiltonian defined by Eq.~(\ref{EnvelopeApprox}) provides the correct ionization probability differential in energy, i.e., the beyond-dipole results obtained with the simplified interaction are in quantitative agreement with the corresponding calculations performed with the full correction term Eq.~(\ref{SchrodH1st}). We would like to remark that, in agreement with previous findings \cite{Forre2005, Forre2014, Toyota2009}, the most substantial difference between the dipole and nondipole results is that in the latter case a larger differential yield in the proximity of the ionization threshold is observed, corresponding to the emission of low-energetic electrons (see the inset in the upper panel of Fig.~\ref{Fig2}). This is due to the nonadiabatic turn-on/off of the laser pulse and the corresponding large spectral
width associated with relatively short pulses.
As the pulse length $T$ increases, the pulse becomes increasingly monochromatic and nonadiabatic effects become of less importance, simply due to the smooth temporal variation of the pulse envelope~\cite{Aldana2001, Emelin2014, Simonsen2015}.
This is consistent from the point of view of the envelope approximation, in which the nondipole interaction is proportional to $f'(t)$.

Figure~\ref{Fig3}~shows the corresponding angular-differential probability distributions for the same laser-atom interactions as shown in Fig.~\ref{Fig2}. The horizontal lines indicate the axis of polarization whereas the vertical black arrows point in the propagation direction of the laser. Here the panels show, in descending order, the results obtained by solving the Schr\"{o}dinger equation with the Hamiltonians~(\ref{SchrodH1st}), (\ref{SchrodHDip}) and (\ref{EnvelopeApprox}), respectively. In the dipole description of the interaction, the electrons are emitted along the axis of
polarization only, whereas in the beyond-dipole case, electrons are also ejected in the counter-propagation direction. The corresponding structure in the angular distribution, which is known as the {\it nondipole lobe} \cite{Forre2006, Forre2007, Zhou2013}, is
explained by a process in which the electron experiences a temporary push in the laser
propagation direction due to the radiation pressure caused by the combined effort of the electric and magnetic fields. Then, when the laser pulse ramps off at the end of the interaction, the (displaced) electron
falls back in the Coulomb potential of the bare nucleus and scatters off it in the backward (counter propagation) direction.
Therefore, a non-adiabatic turn-off of the laser field is expected to play a significant role in the beyond-dipole ionization mechanism. The envelope approximation (lower panel) and the reference Hamiltonian including the
carrier effects (center panel) both give rise to angular distributions which do include the nondipole lobe. Moreover, the results are in fact essentially identical, indicating that, also here, the envelope
approximation accounts for the dominating ionization dynamics beyond the dipole approximation.

Our analysis suggests that the validity of the approximations (\ref{EnvelopeApprox})
and (\ref{DiracH1st}) prevails as the central frequency $\omega$ increases indefinitely. The findings shown in Fig.~\ref{Fig4} support this. Here we have plotted the ionization probability as a function of $\omega$ for the
electric field strength $E_0=m\omega c/(10e)$, and for a pulse duration
corresponding to 10 optical cycles in all cases. As $eE_0/(m \omega)$ is the maximum speed attainable for a free classical electron exposed to the present
laser field, this choice ensures only small relativistic corrections -- if any, while the field remains strong enough to ensure significant corrections to the dipole approximation. Again we see that the approximate Hamiltonian~(\ref{EnvelopeApprox}) provides the same ionization probabilities as the full one, Eq.~(\ref{SchrodH1st}), for all frequencies -- including when the photon energy $\hbar \omega$ approaches the ionization potential at $0.5$~a.u. The same is the case when it comes to the ionization yield obtained with the Dirac equation, with the exception of a small upward shift which is seen in the high $\omega$-regime. This is not surprising considering that the approximation has been implemented in slightly different ways for the relativistic and the non-relativistic case, c.f., Eq.~(\ref{DiracH1st}) and Eq.~(\ref{EnvelopeApprox}), as discussed in Sec.~\ref{SubSectTDDE}.

As the dipole interaction with linearly polarized light only drives transitions between states with the same
azimuthal quantum number ($\Delta m=0$) and our initial state is the
ground state, the population of $m\neq0$-states serves as an overall measure of
the significance of the correction to the dipole approximation. Moreover, the population in different angular channels may be studied directly in the time domain. In Fig.~\ref{Fig5} the population in all $m\neq0$-states is plotted as a function of time both including and excluding the spatial dependence of the carrier. In this case, we have applied a pulse shape given by
\begin{equation}
\label{FermiDiracPuls}
f(t)=\frac{(e^{- \sigma T/2} +1)^2}{(e^{\sigma(t-T/2)}+1)(e^{\sigma(-t-T/2)}+1)} \quad .
\end{equation}
For small $\sigma$ this envelope provides a Gaussian-like pulse shape, c.f., Eq.~(\ref{Gauss}), while it becomes increasingly rectangular in the limit $\sigma \rightarrow \infty$. For intermediate $\sigma$-values it is trapezoidal-like, albeit with a continuous derivative. Figure~\ref{Fig5} shows the total population in $m \ne 0 $ states for three different
choices of $\sigma$. While Eq.~(\ref{EnvelopeApprox}) gives rise to the correct final probabilities at the end of the
interaction, the corresponding wave functions differ from the original ones at intermediate times. Specifically, the full interaction provides an $m\neq0$-population which is highly oscillatory during the laser-atom interaction, while the approximation does not include these oscillations. As such, Eq.~(\ref{EnvelopeApprox}) provides a rather stable frame.

For trapezoidal-like and near-rectangular envelopes, such as the one in the lower panel of Fig.~\ref{Fig5}, the envelope only has a non-vanishing derivative towards the ends.
Thus, any net population shifts induced by interactions beyond the dipole approximation tends to occur during the more or less harsh, non-adiabatic ramp-on and ramp-off of the radiation pressure, as is clearly seen in Fig.~\ref{Fig5} -- in particular in the lower panel. This is consistent with a physical picture in which the radiation pressure in the propagation direction is more or less constant at intermediate times, c.f., Eqs.~(\ref{eq:HamPGFull}) and
(\ref{eq:HamPGEnv}). The nondipole transitions turn out to be larger towards the end of the pulse than at the onset for all the panels in Fig.~\ref{Fig5}, which indicates that the dominating dynamics take place when the radiation pressure suddenly ramps off, i.e., when the displaced electron is pulled back toward the nucleus.

The fact that fluctuations in the populations seen in Fig.~\ref{Fig5} vanish in the envelope approximation could suggest that this approximation is numerically favorable when it comes to the time propagation. Indeed, with the Arnoldi type propagator, a slight speedup is found due to the fact that a smaller dimension of the Krylov sub-spaces is required. However, as the rather rapid dipole dynamics still need to be resolved, this advantage is minor.

A significant numerical advantage is seen, however, when the convergence in terms of partial waves is considered. Figure~\ref{Fig6} shows the photoelectron distribution for energies around the first photoionization resonance following an interaction with a laser pulse of frequency $\omega = 3.5$~a.u. and electric field strength $E_0 = 45$~a.u. Strong oscillatory fringes within the dipole approximation are usually observed in the photoelectron emission spectra (c.f., upper panel Fig.~\ref{Fig6}). These can be traced to the interference between wave packets ejected during the laser ramp-on and ramp-off \cite{Toyota2007,Toyota2008,Demekhin2012}. However, the displacement of the electronic wave packet along the propagation direction, induced by interactions beyond the dipole approximation, partially prohibits this interference resulting in a smoother, less oscillatory distribution. The photoelectron spectrum is thus highly sensitive to the number of partial waves used in our truncated expansion, and, as such, the highest angular momentum number, $l_\mathrm{max}$, used in the expansion provides an ideal quantity to probe. The presence/absence of oscillatory fringes is more pronounced for longer laser exposure times. Therefore we have increased the pulse duration to 40 optical cycles in order to demonstrate a very different convergence behaviors. In the intermediate and lower panels, the results obtained with the Hamiltonians~(\ref{SchrodH1st}) and (\ref{EnvelopeApprox}) are shown for different values of $l_\mathrm{max}$. In the two cases, the shape of the distributions are fully converged at $l_\mathrm{max} = 50$~(exact Hamiltonian) and $l_\mathrm{max} = 20$ (envelope Hamiltonian) indicating significantly better numerical performance in the latter case.
This is mainly due to the fact that the envelope approximation effectively reduces the
maximum strength of the beyond-dipole interaction by one half, which might
seem like a minor advantage considering that the dipole interaction is an order $c$~stronger. However, the asymmetric characteristic of the nondipole field causes a non-zero displacement
of the wave packet, which in turn makes computations more demanding with respect to convergence criteria,
despite the comparably weaker interaction strength.

\section*{Conclusion}
\label{Conclusion}

We have studied the photon-induced break up of the hydrogen atom in a wide range of high-frequency super intense fields by solving the time-dependent Schr\"{o}dinger and Dirac equations including the laser-atom interaction beyond the dipole approximation. We showed that for a large class of examples involving laser ionization, the spatial dependence of the envelope function of the laser pulse provides practically all corrections to the dipole approximation.
Based on the numerical evidence, and a more analytical discussion regarding the correction terms to the dipole approximation, we have proposed a simplified beyond-dipole interaction with a temporal factor dependent only on the laser pulse profile and its time derivative. This simplification provides both an interpretation of the underlying dynamics as well as a numerically favorable scheme. It is seen that any net contribution to the ionization dynamics beyond the dipole approximation tends to occur predominantly at the onset and at the end of the laser pulse. Numerical studies suggest that the latter tends to dominate the former.

\section*{Acknowledgments}

This work was supported financially by the Bergen Research Foundation and the Norwegian Metacenter for Computational Science (Notur), as well as by the Swedish Research
Council, Grant No. 2012-3668.nondipole-induced
We also acknowledge support for our collaboration through the Nordic Institute for Theoretical Physics (Nordita),
the Norden Network NANOCONTROL and the European COST Action XLIC CM1204.
Numerical calculations were carried out at the Cray XE6 (Hexagon) supercomputer installation at Parallab at the University of Bergen and at the Alto cloud computing facility at Oslo and Akershus University College of Applied Sciences.


\end{document}